\documentclass{appolb}
	
\usepackage{amsmath}
\usepackage{amssymb}
\usepackage{graphicx}
\usepackage{xcolor}
\usepackage{cite}
\usepackage{url}

\newcommand{\veps}{\varepsilon}

\begin{document}

%=============================================================================
%=============================================================================
%=============================================================================

\title{
%%%%%%%%%%%%%%%%%%
\vspace{-20mm}
\begin{flushright} \bf IFJPAN-IV-2011-5\\ \end{flushright}
\vspace{5mm}
%%%%%%%%%%%%%%%%%%
Colour coherence of soft gluons in the fully unintegrated NLO 
singlet kernels
\thanks
{This work is supported by the 
Polish Ministry of Science and Higher Education grant 
  No.\ 1289/B/H03/2009/37.
\hfill \\
  Presented by Magdalena Slawinska at the
{\em Cracow Epiphany Conference 2011 - on the First Year of the LHC}, January 10-12, 2011}
\author{M. Slawinska,\\
A. Kusina,  S. Jadach and  M. Skrzypek
\address{Institute of Nuclear Physics PAN,\\
ul. Radzikowskiego 152, 31-342 Krak\'ow, Poland }}
}

\maketitle

\begin{abstract}
{\em Abstract:}
Feynman diagrams with two real partons contributing 
to the next-to-leading-order
singlet gluon-quark DGLAP kernel are analysed. 
The infra-red singularities
of unintegrated  distributions are examined numerically. The analytical formulae
are also given in some cases.
The role of the colour coherence effects is found 
to be crucial for cancellations
of the double- and single-logarithmic infra-red singularities.

\vspace{3mm}
\centerline{\em Submitted To Acta Physica Polonica B}
\end{abstract}

\PACS{12.38.-t, 12.38.Bx, 12.38.Cy}
\vspace{5mm}
%%%%%%%%%%%%%%%%%%%%%%%%%%%
\begin{flushleft}
\bf IFJPAN-IV-2011-5\\
\end{flushleft}
%%%%%%%%%%%%%%%%%%%%%%%%%%%

\newpage

%=============================================================================
\section{Motivation}
%=============================================================================

The study presented here is a part of the development of
the fully exclusive next-to-leading order (NLO) Parton Shower Monte Carlo (MC)
for precision QCD predictions for the LHC experiments,
see \cite{Jadach:2010ew,Jadach:2011kc,Jadach:2011cr}.
DGLAP~\cite{DGLAP} 
evolution of parton distribution functions (PDFs)
is modelled in the Monte Carlo within the unintegrated phase space.
A methodology based on the collinear factorisation theorems in physical gauge
based on refs.  \cite{Ellis:1978sf} and \cite{Curci:1980uw} is used.%
\footnote{ In the complementary approach of 
  refs.~\cite{Ward:2007xc,Joseph:2009rh,Joseph:2010cq}
  soft singularities are resummed first and collinear resummation
  is added next.}
The MC program will simulate {\em  exactly} NLO DGLAP evolution of PDFs by itself,
as opposed to using pretabulated PDFs,
provided by the non-MC programs like QCDNUM~\cite{Botje:2010ay}.
For the construction of such a new NLO parton shower MC program
a new {\em exclusive} (fully unintegrated) 
NLO evolution kernels are required
in order to impose NLO corrections,
by means of reweighting the LO distribution,
as outlined in ref.~\cite{Jadach:2011cr}.
In this method LO MC parton shower 
has to be reconstructed from the scratch,
contrary to methodology of ref.~\cite{Frixione:2002ik},
where at the price of MC weights being non-positive,
one is able to use standard LO parton shower MC.

Very schematically, the corresponding MC weight reads:
\begin{equation}
\text{MC weight} = 
\frac{\text{exact NLO diagram distribution}}{\text{crude LO distribution}}. 
\end{equation}
The potential problem is that a
single Feynman diagram, or small subset of diagrams,
entering the exclusive NLO kernel (and the MC weight)
generally is not gauge-invariant and 
may feature  uncancelled soft singularities.
The Monte Carlo weights may then explode,
unless the {\em crude} distribution of the LO MC already
reproduces exactly soft singularities of NLO diagrams.
It is therefore very important to understand in fine
detail the structure of soft and collinear singularities
in exclusive kernels, as implemented in the LO MC and also including
complete NLO corrections.

For the purpose of the MC we are going to
analyse the example diagrams  and their groups one by one, gaining 
the detailed knowledge about the structure of collinear and soft singularities 
of each Feynman diagram contributing to the NLO kernel
and the interplay between diagrams.
We shall exploit tools and methods of the graphical analysis
of the infra-red singularities which were already
used for the non-singlet diagrams in ref. \cite{Slawinska:2009gn}.
Here, we will extend this study to a gauge invariant subset
of two-real singlet diagrams contributing to the $P_{gq}$ NLO DGLAP kernel.
Let us stress that
the cancellations discussed in the following
are not of the usual KLN \cite{KLN} nature,
i.e. between the real and virtual Feynman diagrams, 
but rather among the real diagrams alone,
and are governed by the spin and colour quantum numbers.
The contributions of the diagrams to the standard DGLAP
(inclusive) kernels  analysed in the following
have been already defined and used
in refs.~\cite{Furmanski:1980cm} and \cite{Ellis:1996nn}.
Generally, we shall examine the structure of the
soft and collinear singularities of the unintegrated distributions
related to these NLO DGLAP evolution kernels.

%=============================================================================
\section{Singlet diagrams considered}
%=============================================================================

The singlet diagrams considered in this contribution originate from the LO amplitude
for splitting a gluon into a quark (antiquark)
\begin{center}
\includegraphics[width=0.8cm, height=1cm]{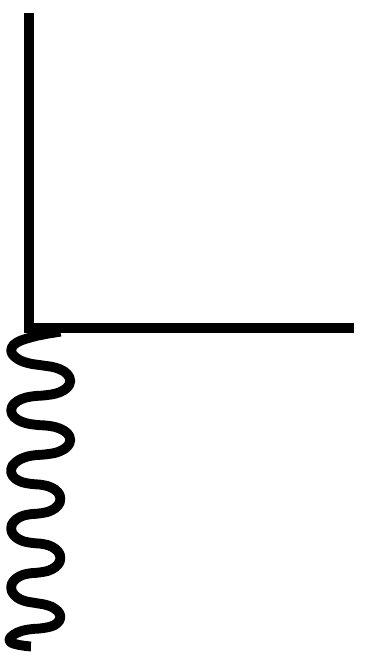}
\end{center}
by means of adding the  NLO corrections from the emission of an additional gluon:
\begin{figure}[h!]
\center
\raisebox{-10pt}{\includegraphics[width=0.8cm, height=1cm]{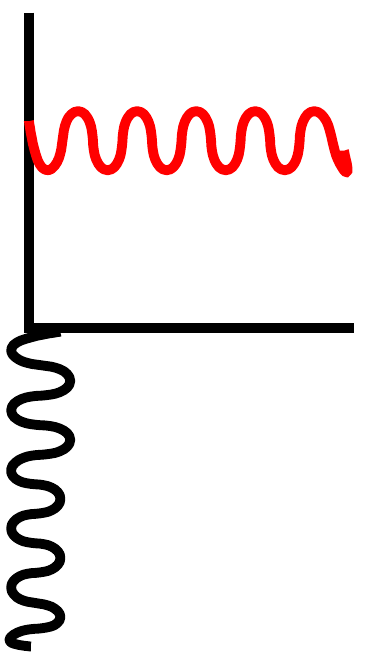}}\quad \quad 
+ \quad \quad
\raisebox{-10pt}{\includegraphics[width=0.8cm, height=1cm]{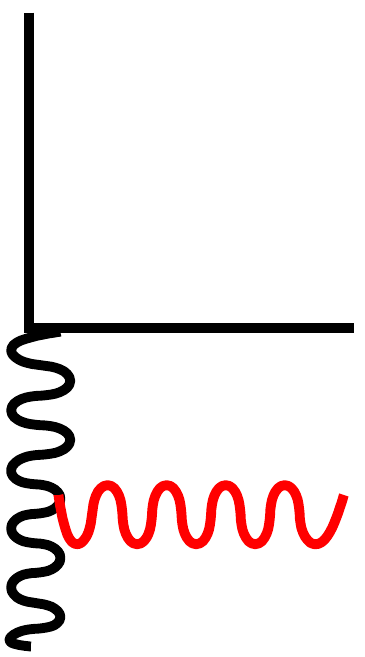}}\quad \quad
+ \quad \quad
\raisebox{-10pt}{\includegraphics[width=0.8cm, height=1cm]{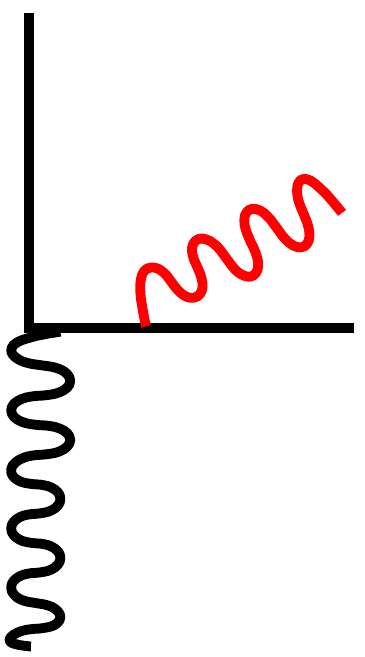}}
\end{figure}

\begin{figure}[h]
\center
\includegraphics[width=7cm]{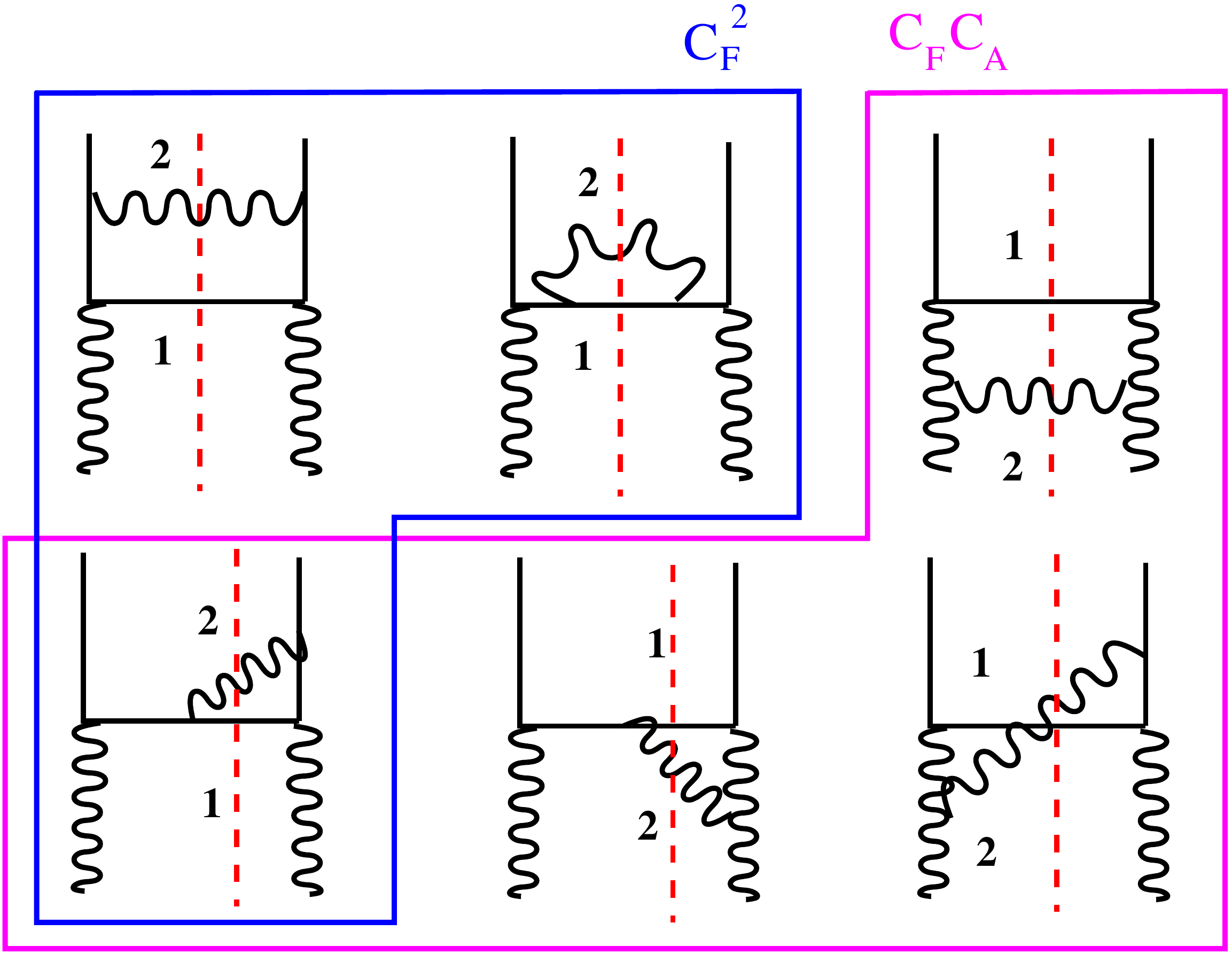}
\caption{Singlet gluon-quark diagrams; "1" denotes a  quark and "2" - a gluon.}
\label{fig:singlet}
\end{figure}
Feynman diagrams contributing to the NLO kernel result
from squaring the above sum of amplitudes and
are displayed in Fig.~\ref{fig:singlet}.
It is worth noting, that the ladder diagrams 
(the first and the third in the upper 
row in Fig.~\ref{fig:singlet}) enter the NLO kernel
supplemented with the so called collinear counterterms 
that subtract off the leading-order contributions.
In this contribution, however,
the counterterms will be included only at the end of the analysis,
and before that the leading-order singularities will be visible. 

We will adopt the same approach as in ref. \cite{Slawinska:2009gn}
and analyse all distributions in the logarithmic Sudakov variables 
$(\ln(a_1/a_2), \ln(\alpha_1/\alpha_2))$,
where  $\alpha_i$ come from the Sudakov
parametrisation of four-momenta of the emitted particles: 
$k_i =  \alpha_i p + \alpha_i^- n + k_{i\perp}$ and $a_i$ are angular (rapidity-related) variables,
$ a_i = \frac{|{\bf k}_{i\perp}|}{\alpha_i}$, $i=1, 2$.
All contributions are normalised to the eikonal phase space:
\begin{equation}
d\Psi = \frac{d\alpha_1}{\alpha_1}
        \frac{d\alpha_2}{\alpha_2}
        \frac{d a_1}{a_1}
        \frac{d a_2}{a_2},
\end{equation}
with the angles integrated over and terms of order 
$\mathcal{O}(\veps)$ neglected.
Moreover, we will ensure that at least one emission is hard by constraining 
$\alpha_1 + \alpha_2 = 1-x > 0$. 
Similarly, the maximal angle is fixed to an arbitrary
parameter. 
Hence, the ratios $a_1/a_2$ and $\alpha_1/\alpha_2$ 
will measure  the relative
hardness and angles of the two partons.

We will explore the soft limit of the diagrams in Fig. \ref{fig:singlet}, namely
the limit where both $|{\bf k}_{i\perp}|\rightarrow 0$
and $\alpha_i \rightarrow 0$ (for a given $i$), but $a_i$ remains finite. 
In the logarithmic Sudakov  variables the singularities will appear on the plots as one- or two-dimensional infinite structures.

\section{Results}

Let us first consider the $C_F^2$ class of diagrams, corresponding to 
emission of a gluon from a quark.
\begin{figure}
\begin{displaymath}
\begin{split}
{\includegraphics[height=1.2cm]{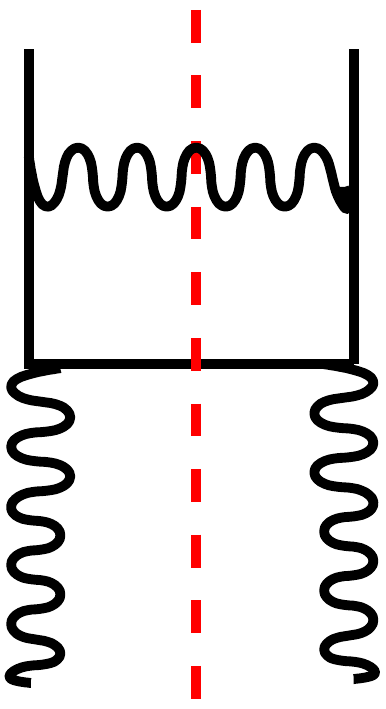}}
\hspace{2cm}\raisebox{10pt}{+}&\hspace{2cm}
{\includegraphics[height=1.2cm]{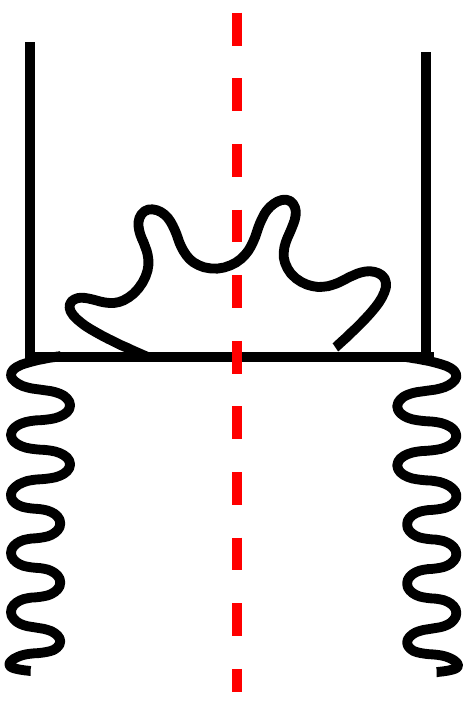}} 
\hspace{2cm}\raisebox{10pt}{=}\hspace{2cm}
\\
\raisebox{-30pt}{\includegraphics[width=3.8cm]{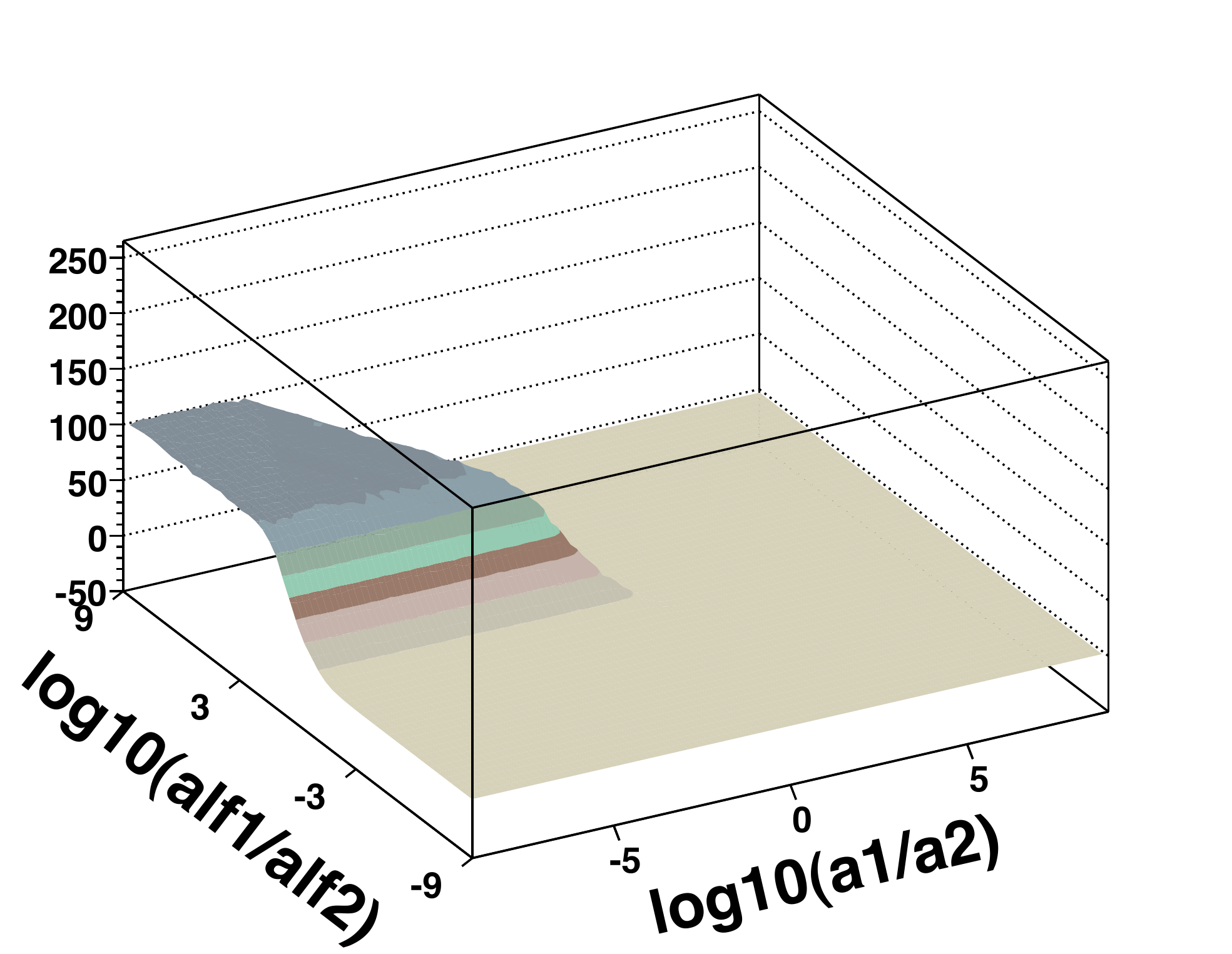}}
\raisebox{15pt}{\Large +} &
\raisebox{-30pt}{\includegraphics[width=3.8cm]{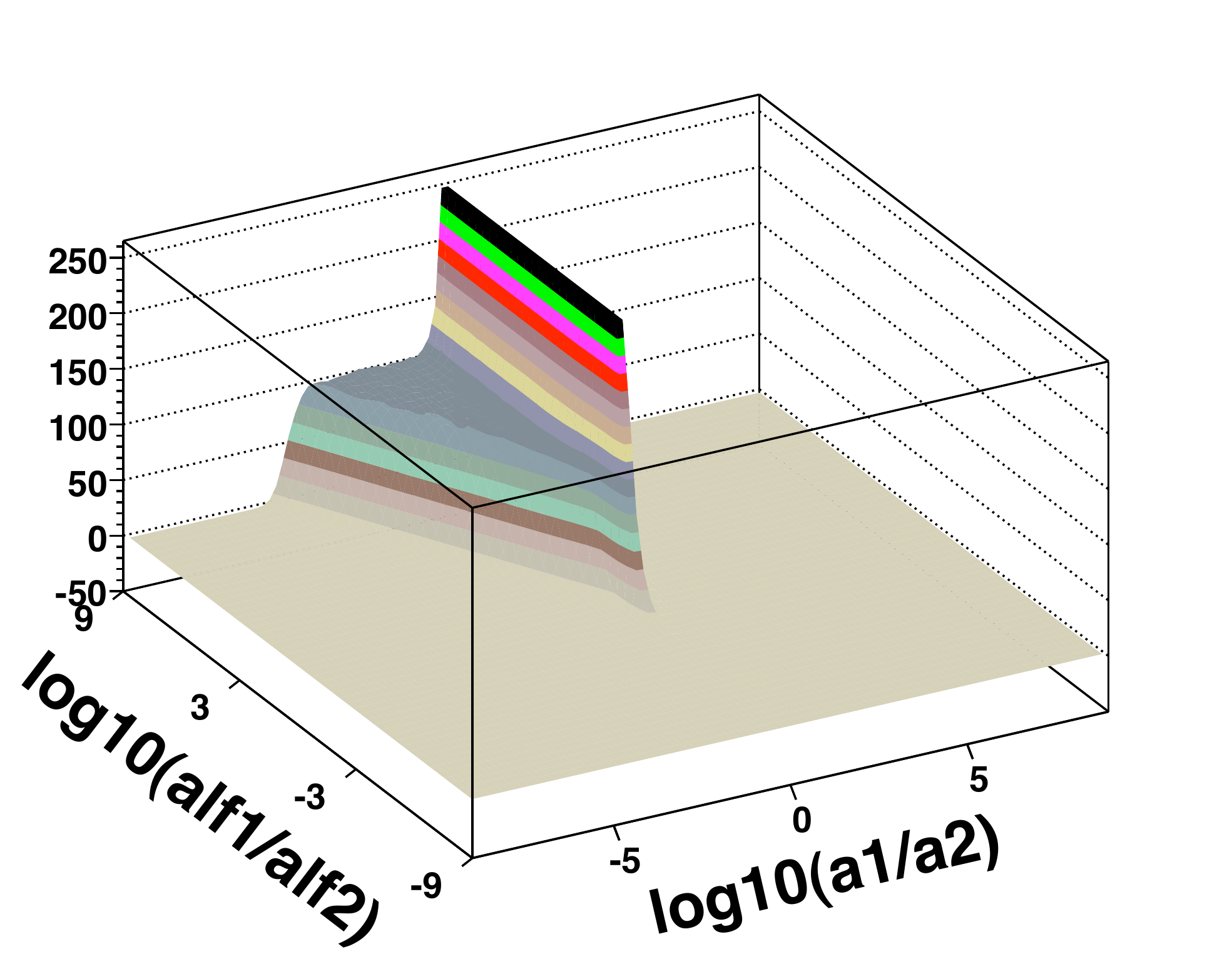}} 
 \raisebox{15pt}{\Large = }
\raisebox{-30pt}{\includegraphics[width=3.8cm]{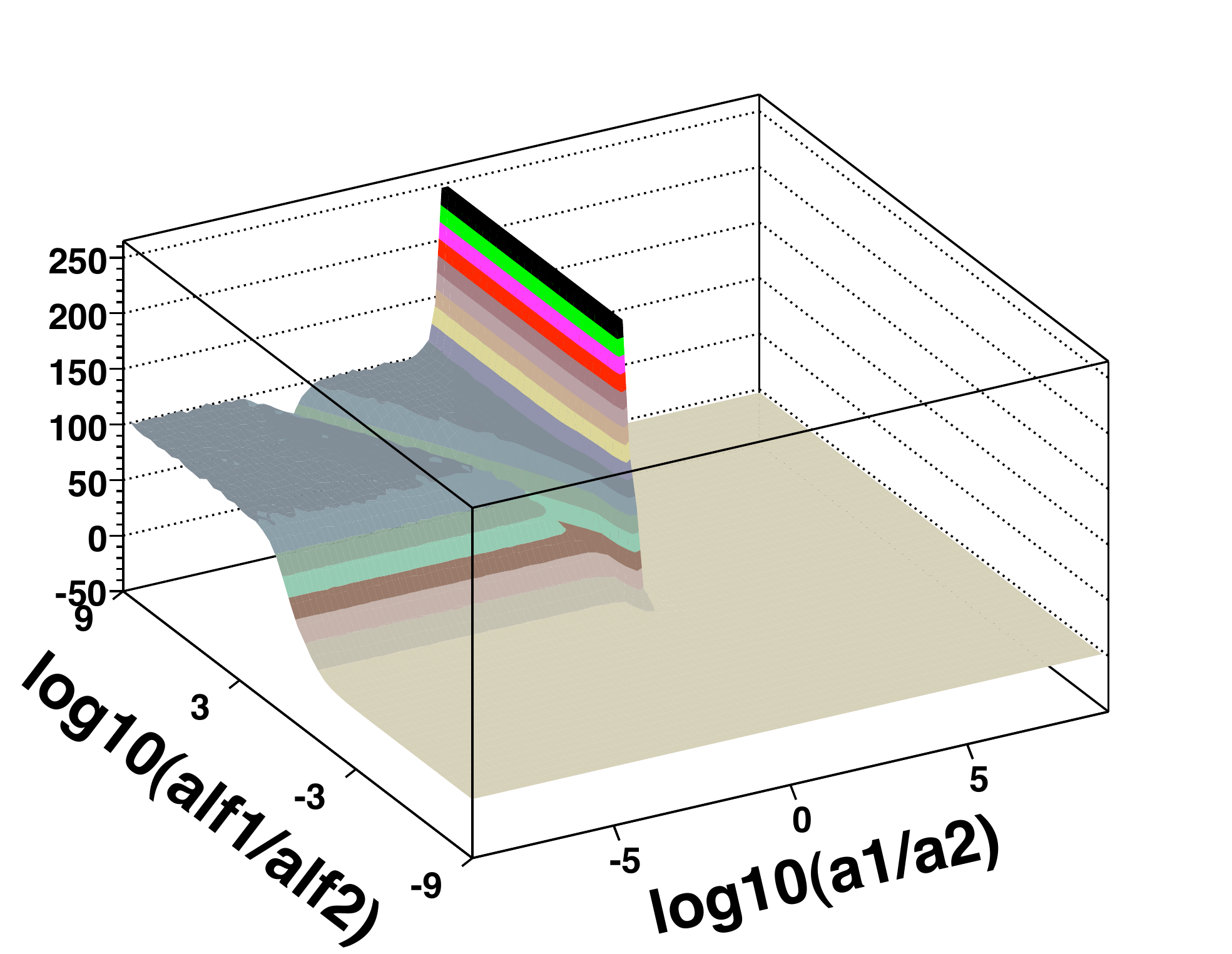}}
\end{split}
 \end{displaymath}
 \caption{Amplitude-squared diagrams $\sim C_F^2$.}
\label{fig:aCF2}
 \end{figure}
% \vspace{-4.2cm}
The bremsstrahlung type diagram is displayed in Fig. \ref{fig:aCF2} (left). 
It has a doubly-logarithmic singularity visible as the
infinite trapezoidal plateau, bordered by the lines $\alpha_1=\alpha_2$ and
$\frac{a_2^2}{a_1^2}  = \frac{\alpha_1}{\alpha_2}$ (the line of equal 
``lightcone minus variables'' $\alpha_i^-$). 
The second diagram, representing the amplitude-squared of the emission of a gluon from
the emitted quark (Fig.~\ref{fig:aCF2}, middle), 
features a collinear singularity manifesting itself as the infinite ridge along
the line of equal  angles. 
This singularity, however, is not related to the soft limit
(it is compensated by the virtual diagram) 
and will not be considered here. 
This diagram has also a doubly-logarithmic singularity 
in the form of a triangular plateau,
bordered by the lines of equal angles and ``minus variables''. 
The sum of the two (the rightmost plot in Fig.~\ref{fig:aCF2}) 
features two equal-height plateaux with the canyon
at the line of equal minus variables. 

The most singular terms from the distributions of the diagrams 
are necessarily proportional to the products of the leading-order 
DGLAP kernels:
\begin{equation}
\raisebox{-15pt}{\includegraphics[height=1.2cm]{xHgq_cut.pdf}}
\approx C_F^2 \frac{(\alpha_1^2 + (1-\alpha_1)^2)(x^2 + (1-\alpha_1)^2) }{\alpha_1\alpha_2^2}
	\frac{a_2^4}{q^4(a_1, a_2)}
\label{eq:Hgq}
\end{equation}
and 
\begin{equation}
P_{qg}(z_1)P_{qq}(z_2) = C_F^2\frac{\alpha_1^2 + (1-\alpha_1)^2}{2}
		\frac{x^2 + (1-\alpha_1)^2}{(1-\alpha_1)\alpha_2},
\end{equation}
where $z_1=1-\alpha_1$ and $z_2= \frac{x}{1-\alpha_1}$.
Similarly:
\begin{equation}
\raisebox{-15pt}{\includegraphics[height=1.2cm]{xHv_cut.pdf}}
\approx 2C_F^2 \frac{\alpha_1(x^2 + (1-x)^2)}{\alpha_2^2}
\frac{a_1^4 a_2^2}{a^2}\frac{1}{q^4(a_1, a_2)},
\label{eq:Hv}
\end{equation}
\begin{equation}
\begin{split}
P_{qg}(z_1)P_{qq}(z_2) &= C_F^2\frac{x^2 + (1-x)^2}{2}
	\frac{(1-x)^2 + \alpha_1^2}{(1-x)\alpha_2}\\
& _{\overrightarrow{\alpha_2 \rightarrow 0}}\; 
C_F^2 (x^2 + (1-x)^2) \frac{\alpha_1}{\alpha_2},
\end{split}
\end{equation}
where $z_1=1-x$ and $z_2=\frac{\alpha_1}{1-x}$.

In eqs.~\eqref{eq:Hgq} and~\eqref{eq:Hv} we also used 
$ q^2(a_1, a_2) = \frac{1-\alpha_2}{\alpha_2} a_1^2 + \frac{1-\alpha_1}{\alpha_1} a_2^2 + 2 a_1 a_2\cos\phi$
(proportional to the denominator of the propagator of the most virtual quark)
and $a^2 = a_1^2 + a_2^2 - 2 a_1 a_2\cos\phi$ 
(proportional to the invariant mass of the emitted quark and gluon).

\begin{figure}
\begin{displaymath}
\begin{split}
{\includegraphics[height=1.2cm]{xHgq_cut.pdf}}
\raisebox{10pt}{+}
{\includegraphics[height=1.2cm]{xHv_cut.pdf}} 
\hspace{0.5cm}& \raisebox{10pt}{+\hspace{1cm} \hspace{0.5cm}\Large{2}}\hspace{0.5cm}
{\includegraphics[height=1.2cm]{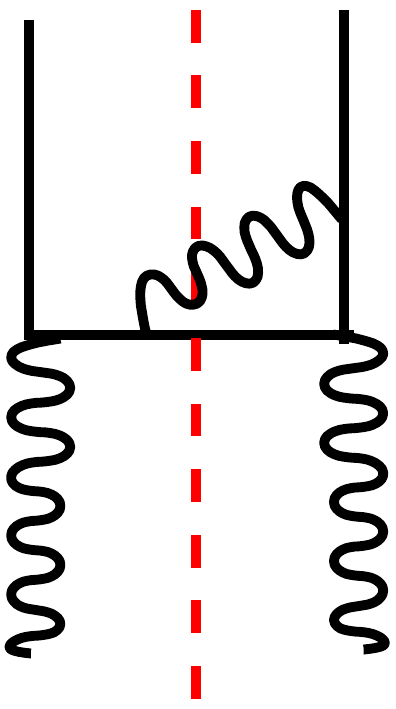}} 
\raisebox{10pt}{\hspace{0.5cm} = }\\
\raisebox{-15pt}{\includegraphics[width=3.8cm]{mcHgq+Hv.pdf}}
& \raisebox{15pt}{\Large +} 
\raisebox{15pt}{\Large 2} \hspace{0.1cm}
\raisebox{-15pt}{\includegraphics[width=3.8cm]{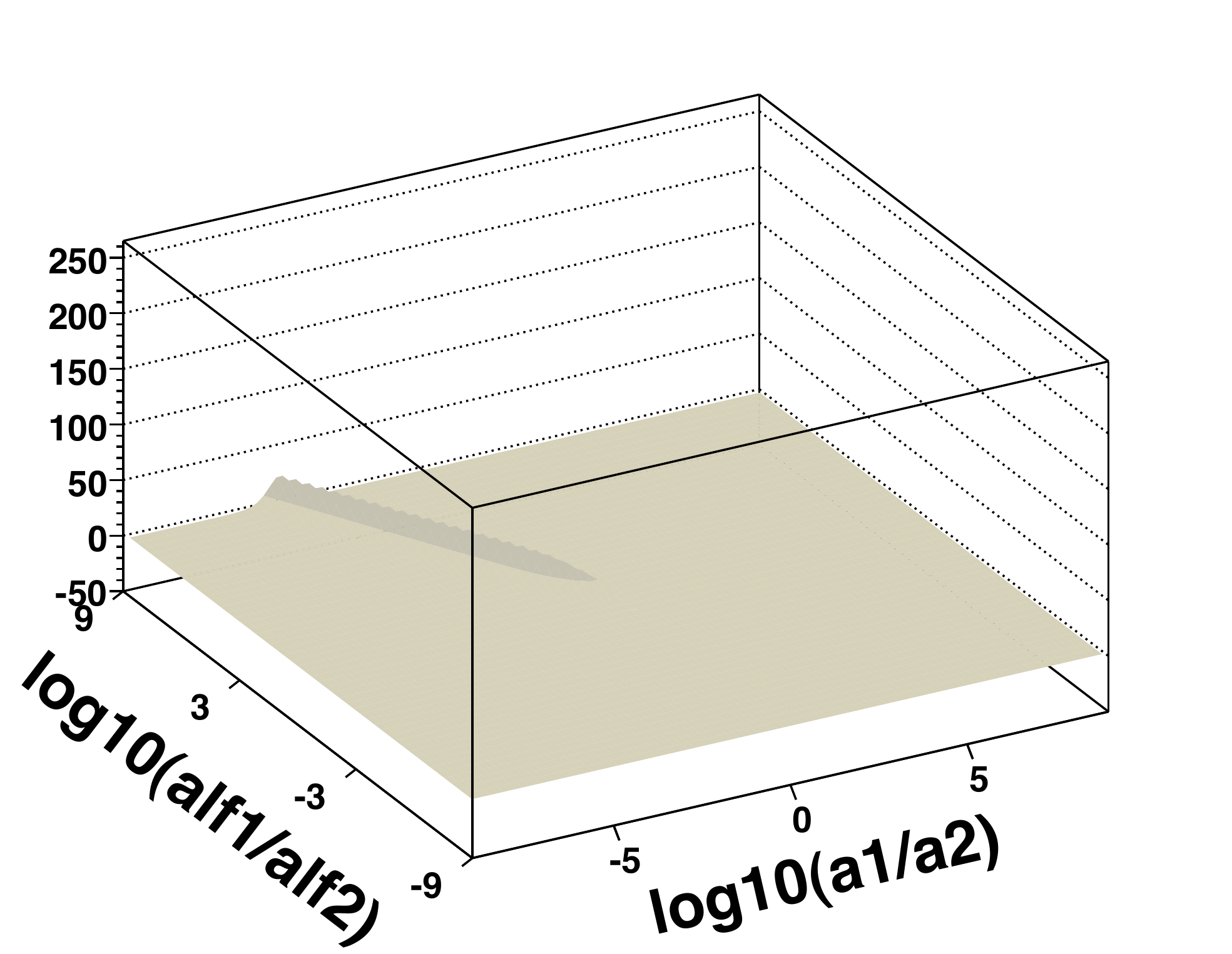}}
 \raisebox{15pt}{\Large = } 
 \raisebox{-15pt}{\includegraphics[width=3.8cm]{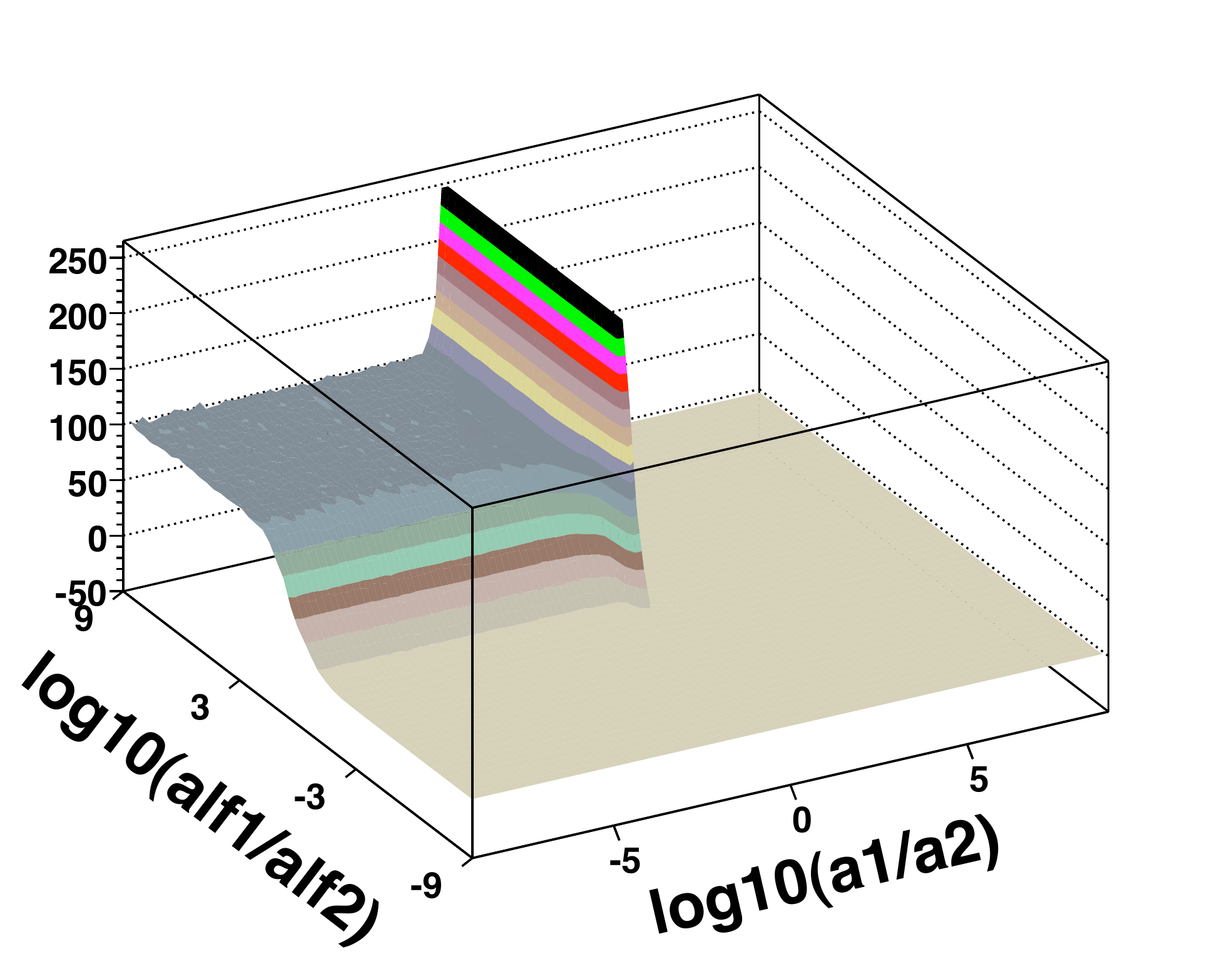}}
\end{split}
\end{displaymath}
\caption{The infra-red cancellations among the diagrams $\sim C_F^2$.}
\label{fig:a+vCF2}
\end{figure}

The ``canyon'' structure in the plot, being the remaining
singly logarithmic singularity, however, spoils the soft limit
regardless of the counterterm employed.
If we add now the interference diagram,%
\footnote{This diagram's colour
 coefficient is equal to $C_F^2 - C_AC_F/2$. In this analysis we add only its 
 part $\sim C_F^2$, adding the other one $\sim - C_AC_F/2$ to the $C_AC_F$
diagrams.}
as shown in Fig. \ref{fig:a+vCF2}, the canyon gets removed.

What remains is the uniform plateau bordered by $a_1=a_2$ 
collinear singularity. 
The ``minus variable'' ordering, 
preferred by each diagram separately,
turns out to be irrelevant for the sum of diagrams!

From the above formulae it follows that 
the quadratic plateau represents the leading-order contribution.
In the NLO kernel
it is removed by the counterterm of the factorisation procedure
proportional to
\begin{equation}
C^{C_F^2} \approx C_F^2
\frac{2(x^2 + (1-\alpha_1)^2)}{1-\alpha_1}
\frac{(\alpha_1^2 + (1-\alpha_1)^2)\alpha_1}{1- \alpha_1} =
4 P_{qg}(z_1)P_{qq}(z_2) \frac{\alpha_1 \alpha_2}{1-\alpha_1}.
\label{eq:3qgcounterterm}
\end{equation}
The doubly logarithmic singularity of the counterterm
depends on the details of factorisation procedure in use.%
\footnote{ 
The counterterm in eq.~\eqref{eq:3qgcounterterm}
has the additional theta function
related to ordering of the emissions (not shown explicitly),
defining the boundaries of the LO plateaux.
In the following we will use the ordering in the angular variables $a_i$.
}

\begin{figure}[h]
\begin{displaymath}
\begin{split}
{\includegraphics[height=1.2cm]{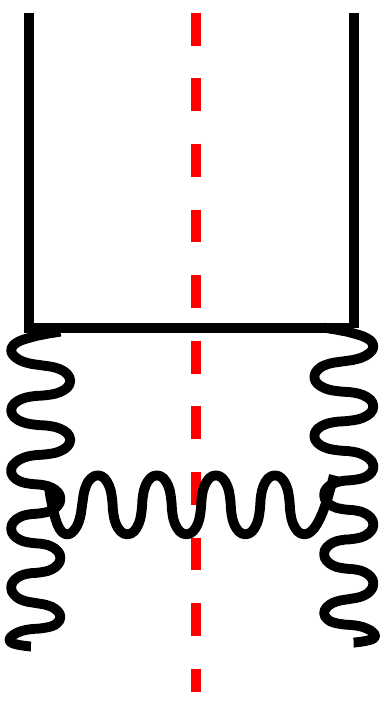}}
\hspace{1cm} & \raisebox{10pt}{+}\hspace{1cm}
\raisebox{10pt}{\Large{2}} \hspace{0.5cm}
{\includegraphics[height=1.2cm]{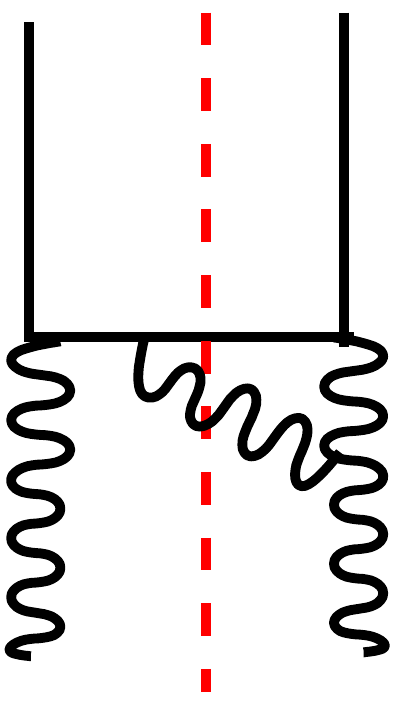}} 
\hspace{1cm} \raisebox{10pt}{=}
\\
\raisebox{-15pt}{\includegraphics[width=3.8cm]{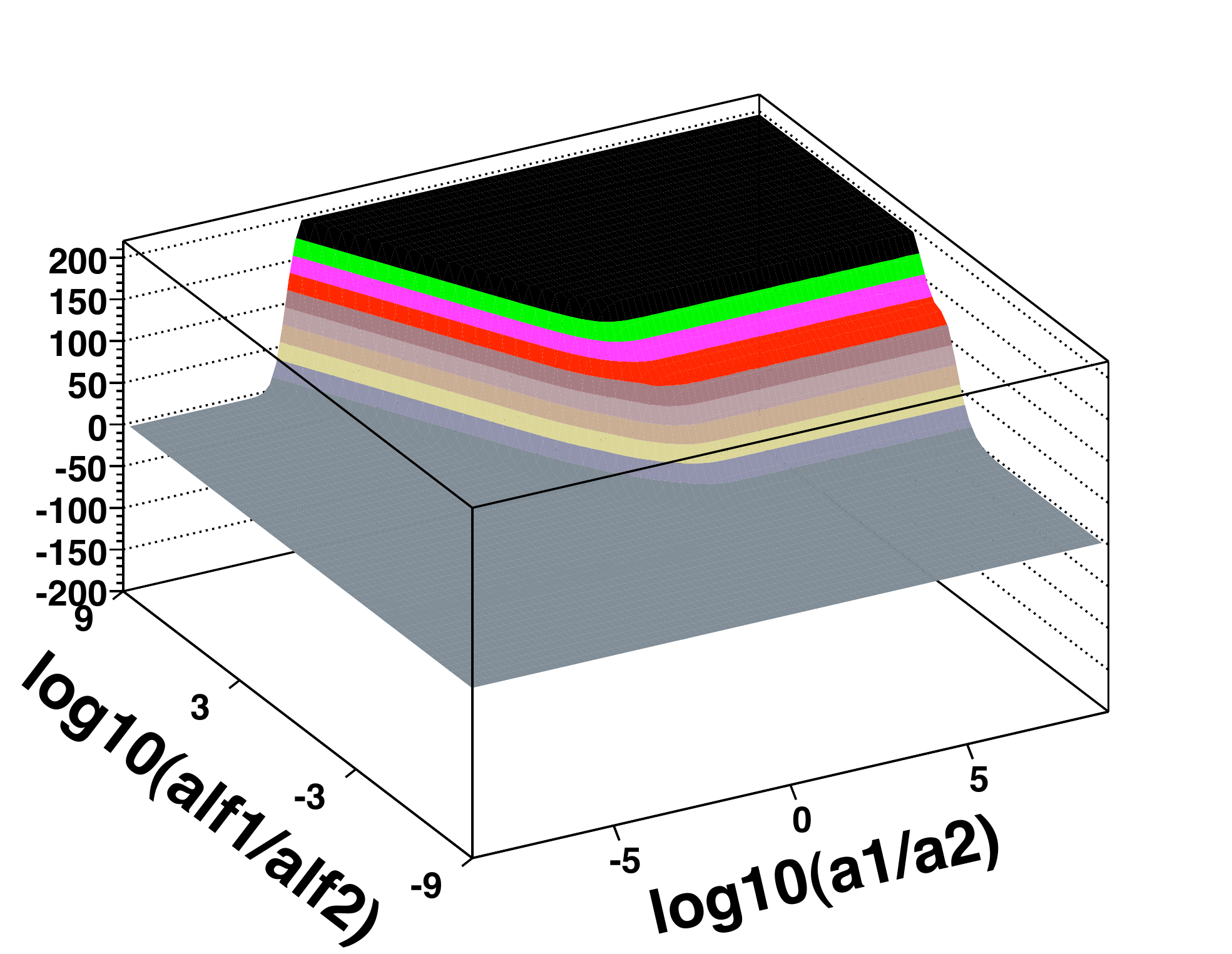}}
& \raisebox{15pt}{\Large + }
\hspace{0.1cm}\raisebox{15pt}{\Large{2} }\hspace{0.1cm}
\raisebox{-15pt}{\includegraphics[width=3.8cm]{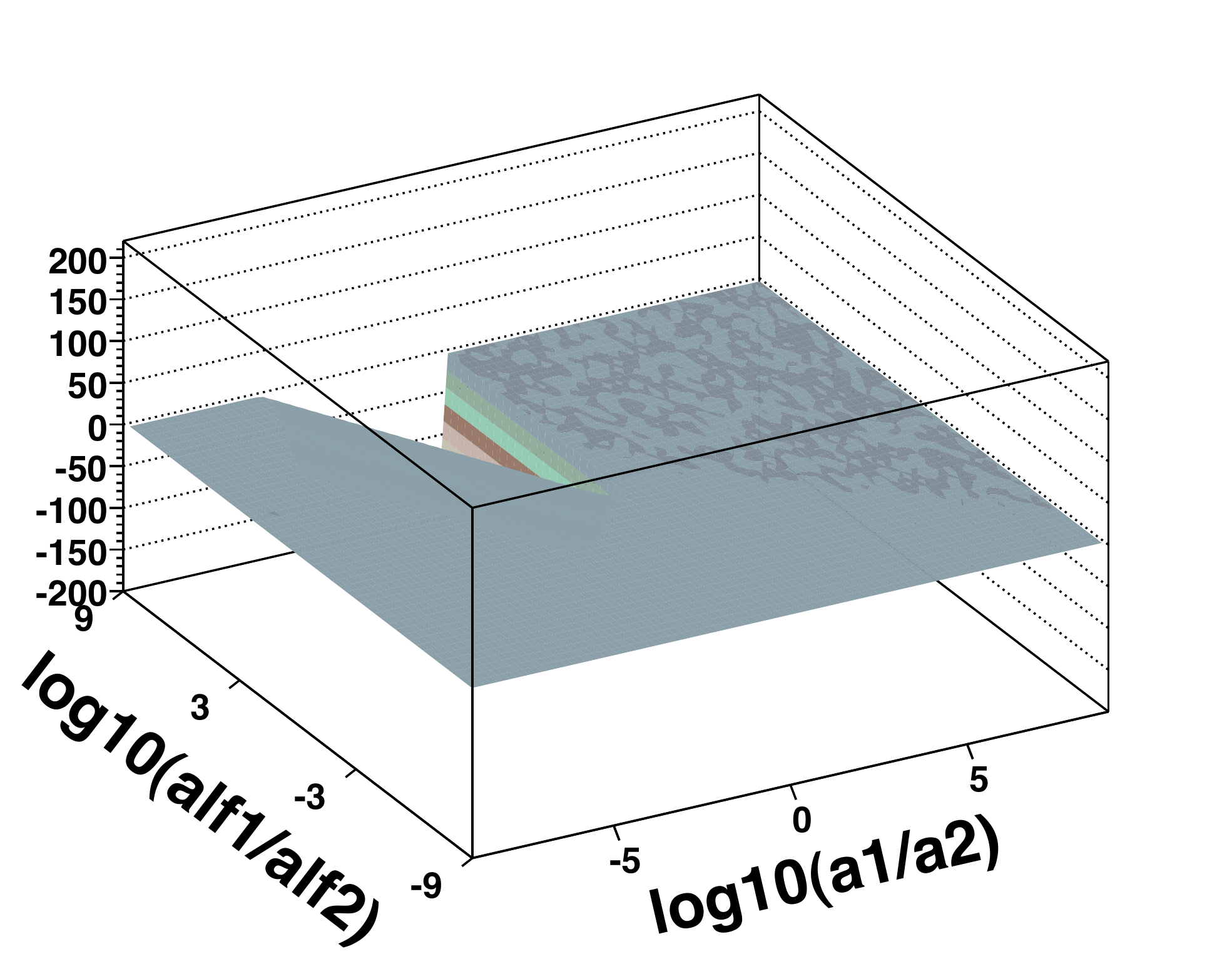}}
 \raisebox{15pt}{\Large = } 
 \raisebox{-15pt}{\includegraphics[width=3.8cm]{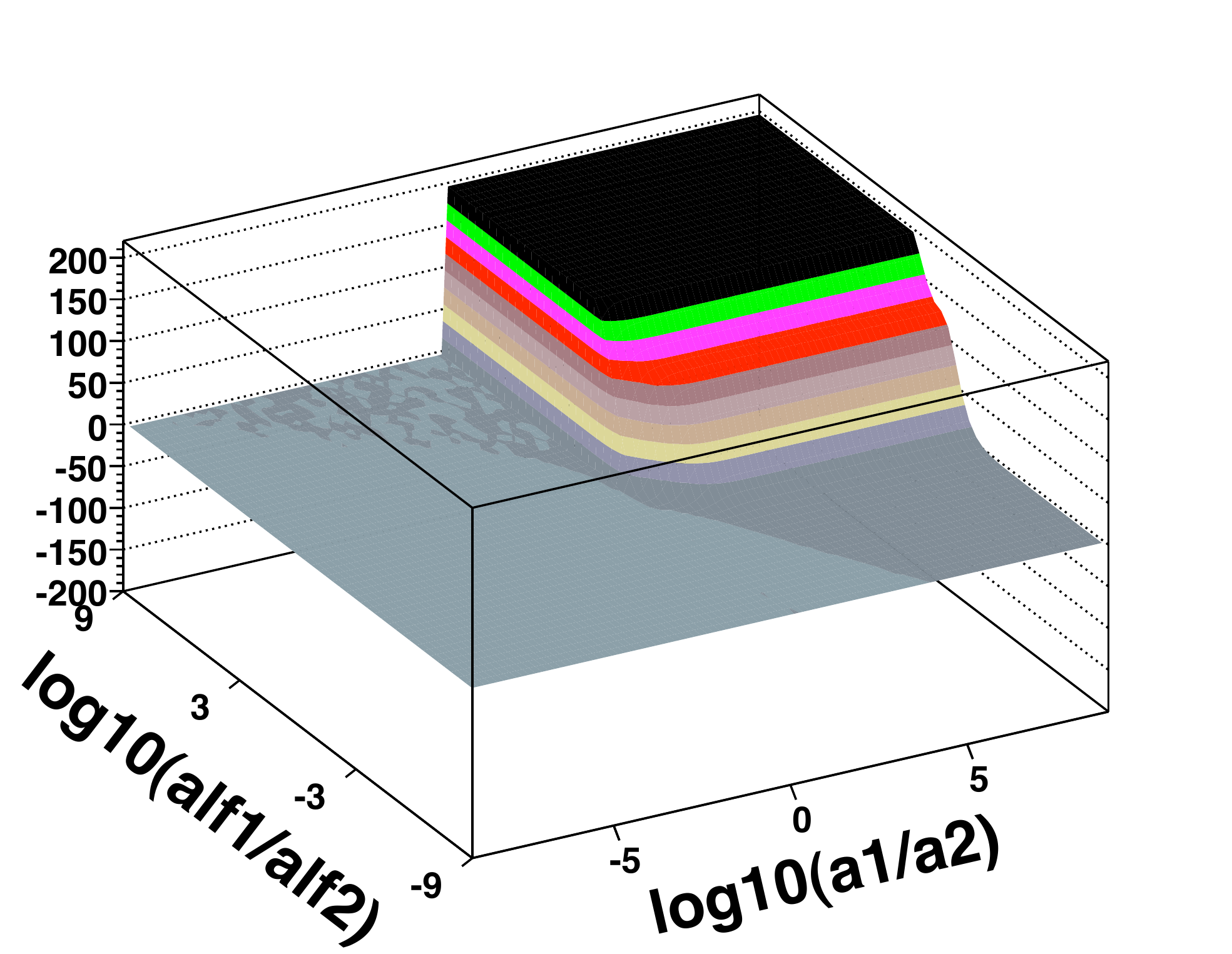}}
\end{split}
\end{displaymath}
\caption{The cancellation of doubly-logarithmic 
singularities among diagrams $\sim C_FC_A$.}
\label{fig:2lCFCA}
\end{figure}

The $C_AC_F$ subset consists of diagrams that correspond to the emission of a 
gluon from the incoming gluon. 
They include only one amplitude-squared diagram. 
Displayed in Fig. \ref{fig:2lCFCA}, left,
it has a doubly-logarithmic singularity 
- the plateau stretching  in the two regions:
where the  angle of the emitted gluon is larger 
than the quark's and vice versa. 
The diagram contributes
in the region of phase space where both emissions are ordered
in the minus variable.
After adding the interference diagram,
the  boundaries of the resulting plateau are corrected
and the sum contributes in the region
of the phase-space, where the angle of the quark is larger than
the angle of the gluon.

The collinear counterterm is given by:
\begin{equation}
C^{C_A C_F} \approx C_A C_F 
\frac{4\alpha_1(\alpha_1^2 + x^2)}{(1-\alpha_2)^2}
\frac{(\alpha_2^2 - \alpha_2 + 1)^2}{(1-\alpha_2)^2}
= 8 P_{gg}(z_1)P_{qg}(z_2)\frac{\alpha_1\alpha_2}{1-\alpha_2}.
\label{eq:q3gcounterterm}
\end{equation}

While the $C_F^2$ counterterm
of eq.~\eqref{eq:3qgcounterterm} features the $a_1 <a_2$ ordering, 
the ordering in the $C_AC_F$ counterterm of eq.~\eqref{eq:q3gcounterterm} 
is the opposite ($a_2 < a_1$) due to a gluon being emitted before a quark. 

\begin{figure}
\begin{displaymath}
\begin{split}
\raisebox{10pt}{\Large{2}} \hspace{0.5cm}
{\includegraphics[height=1.2cm]{xHYf_cut.pdf}}
\hspace{1cm}&\raisebox{10pt}{+}\hspace{1cm}
\hspace{0.5cm}\raisebox{10pt}{\Large{2}} \hspace{0.5cm}
{\includegraphics[height=1.2cm]{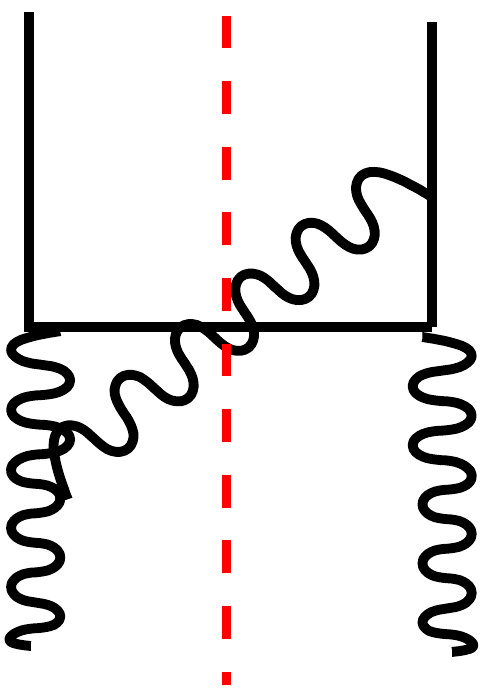}}
\hspace{1cm}\raisebox{10pt}{=}
 \\
\raisebox{15pt}{\Large 2}
\raisebox{-15pt}{\includegraphics[width=3.8cm]{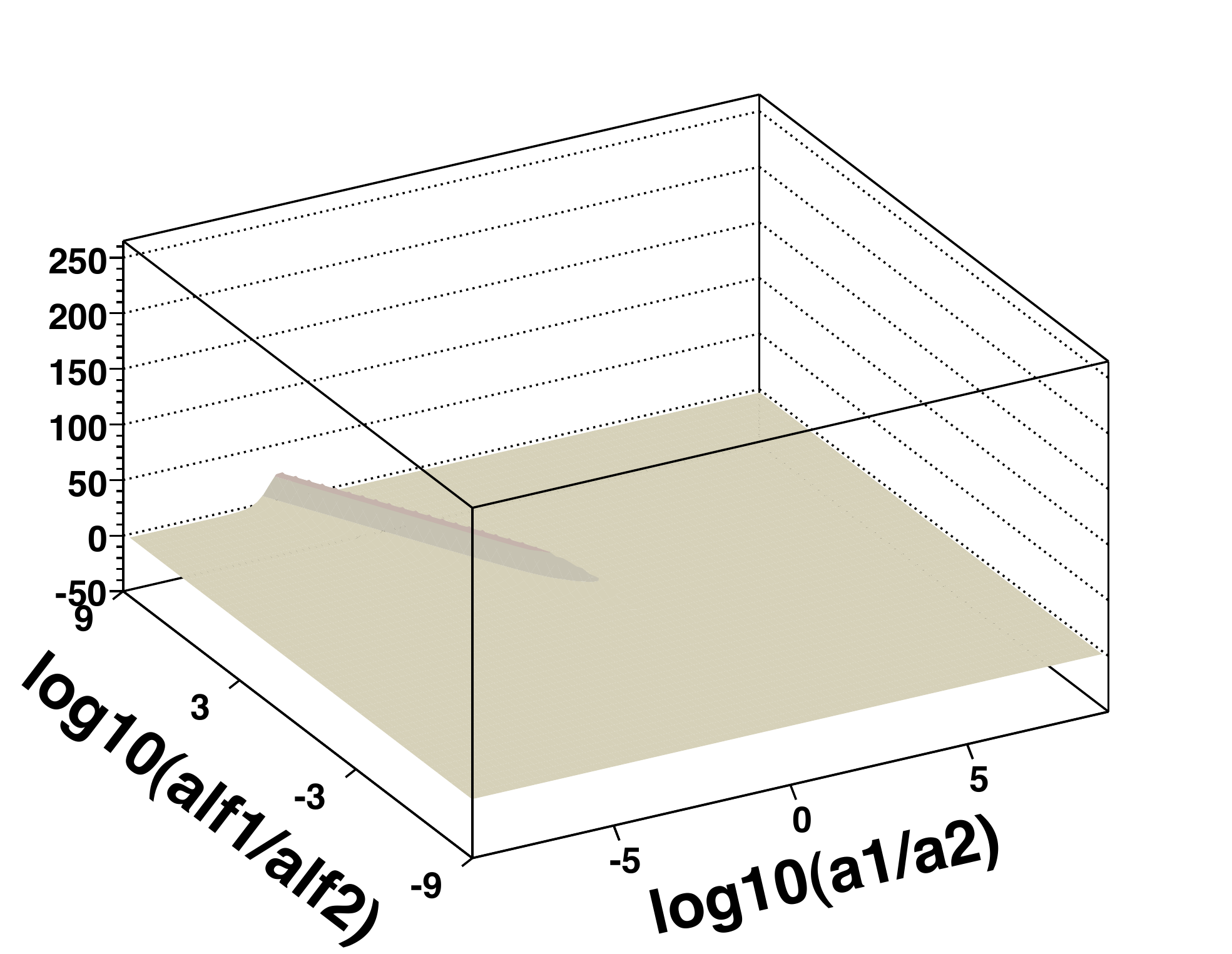}} &
\raisebox{15pt}{ + }
\raisebox{15pt}{\Large 2} 
\raisebox{-15pt}{\includegraphics[width=3.8cm]{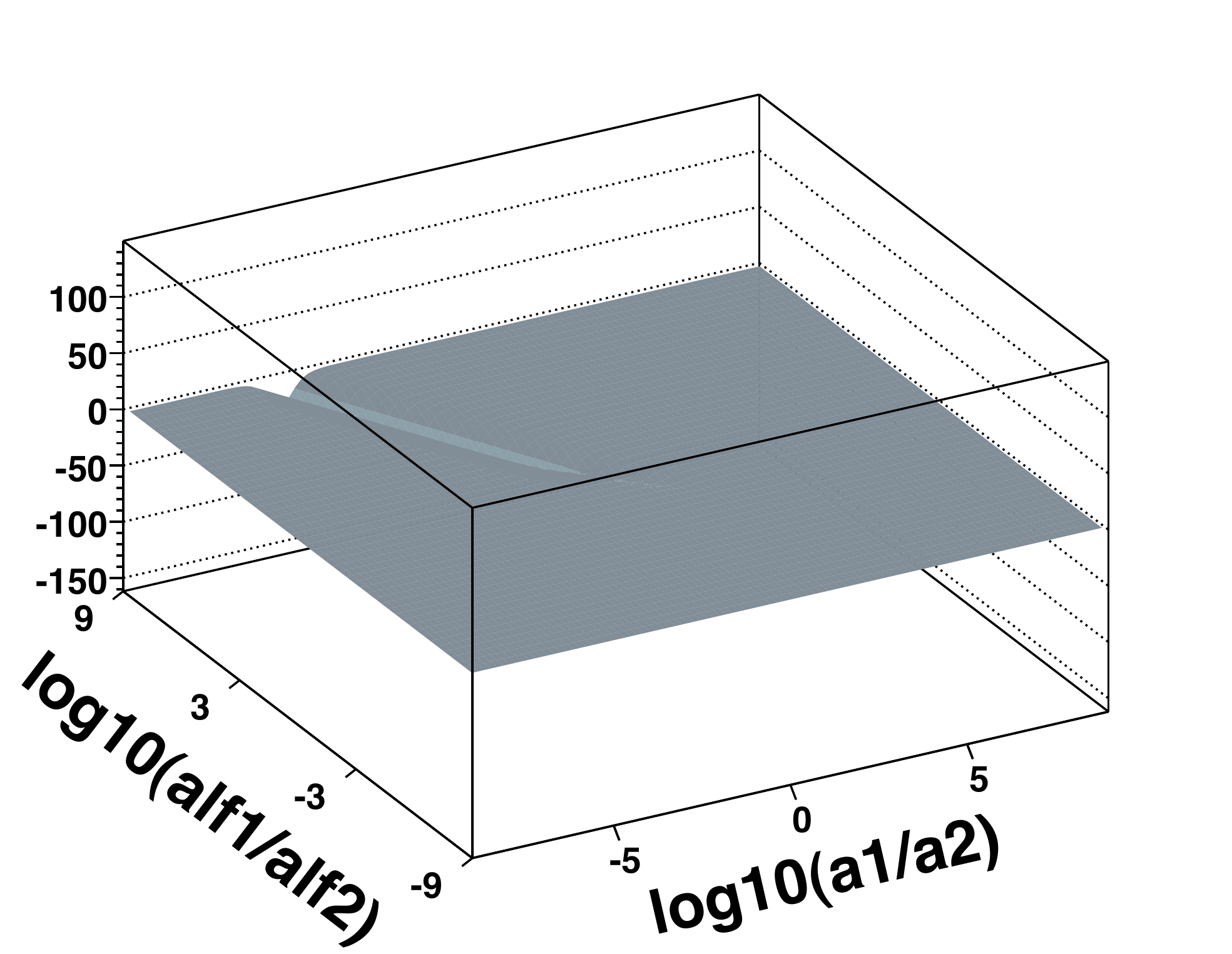}}
\raisebox{15pt}{ = }
\raisebox{-15pt}{\includegraphics[width=3.8cm]{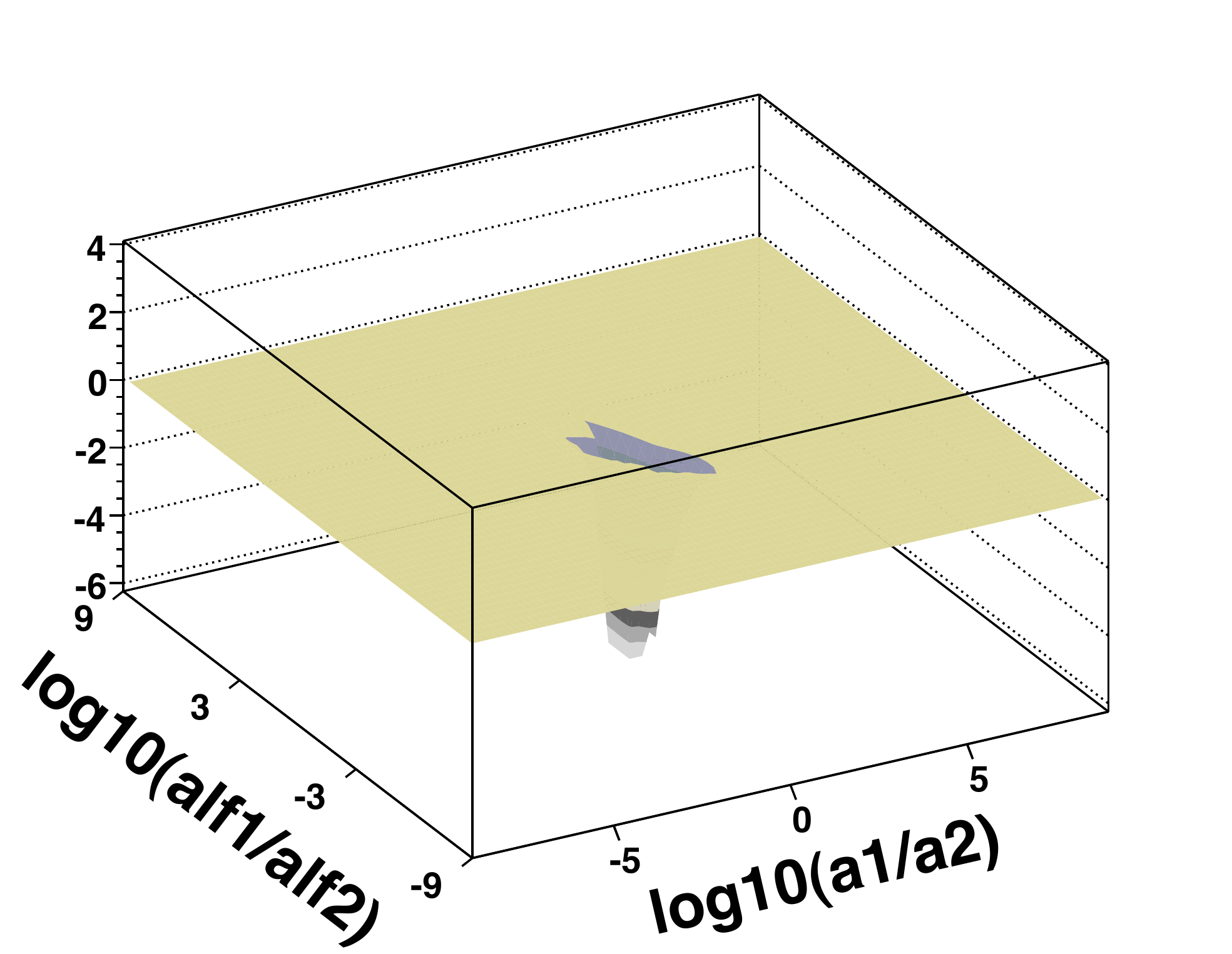}}
\end{split}
\end{displaymath}
\caption{The cancellation of singly-logarithmic 
singularities among interference diagrams $\sim C_FC_A$.}
\label{fig:1lCFCA}
\end{figure}

The remaining $C_AC_F$ interferences feature single-log singularities, 
seen as infinite canyons/ridges along the  line of equal minus-variables 
in Fig. \ref{fig:1lCFCA}  that cancell out when added.
What remains is  the little hill in the central 
region, which leads to a finite contribution.

\begin{figure}[ht!]
\center
\includegraphics[width=6cm]{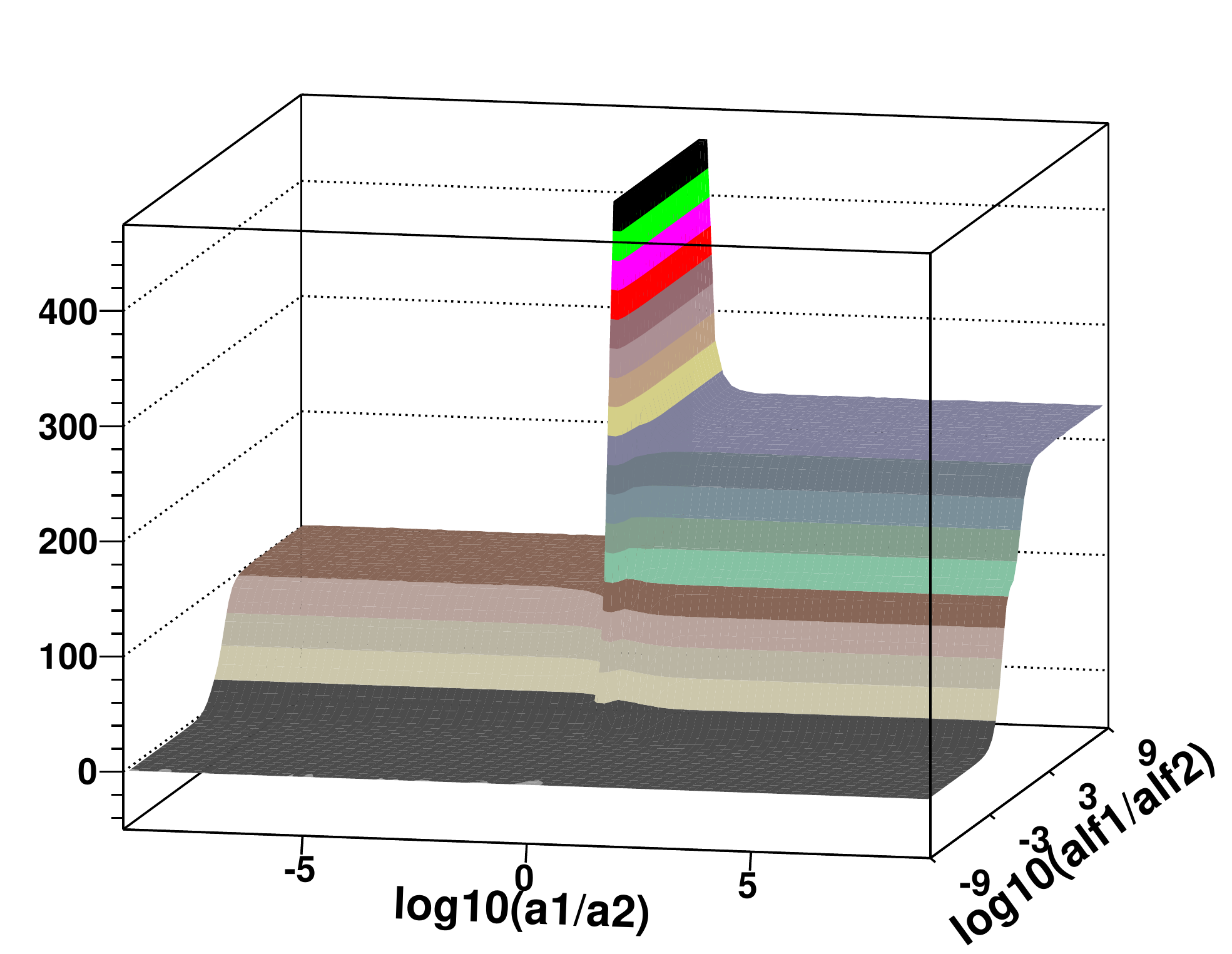}
\includegraphics[width=6cm]{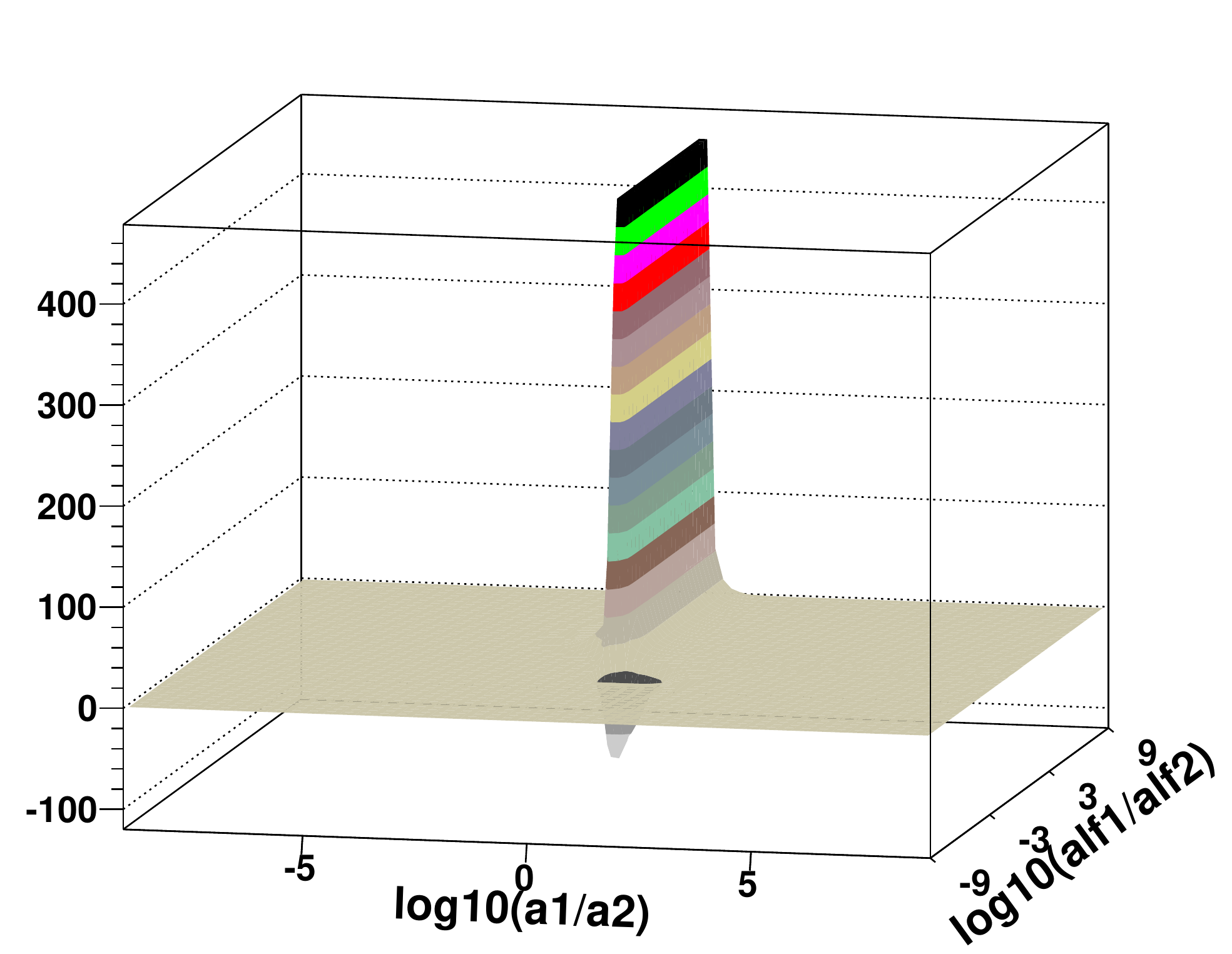}
\caption{All singlet gluon-quark contributions to the NLO kernel
 added together (left) and with counterterms subtracted (right).}
\label{fig:allplot}
\end{figure}

The sum of all singlet diagrams discussed in this contribution 
is presented in Fig.~\ref{fig:allplot}.
The left-hand-side plot in this figure shows two leading-order 
plateaux separated by the line of equal angles
 $a_1 = a_2$. 
The line represents a collinear singularity and comes from the diagram,
in which the additional gluon is emitted from the emitted quark.
The plateau on the left (in brown) corresponds to the topologies in which a
soft gluon is emitted from a quark. 
The right-hand-side plateau (navy-blue)
represents contributions with a soft gluon emitted from the incoming gluon.
The relative height of both plateaux is equal to $ C_F^2/C_FC_A$,
as expected.

In the right plot the same sum is presented, but with the leading order 
singularities cancelled out by the factorisation counterterms
of eqs.~\eqref{eq:3qgcounterterm} and~\eqref{eq:q3gcounterterm}
on the left- and right-hand-side of this plot, respectively.
The plot features the collinear singularity only.

%=============================================================================
\section{Conclusions}
%=============================================================================

We conclude that the restoration of gauge invariance  (colour coherence) is crucial
in cancelling infra-red singularities. 
We understand the soft limits of NLO exclusive kernels, 
observe and explain the cancellations of double- and single-logarithmic
soft singularities.
The angular ordering is the preferred parametrisation of the phase space
in view of the soft singularity
structure of the distributions
from gauge-invariant subset of diagrams contributing to NLO evolution
kernels in the exclusive (unintegrated) form.

%=============================================================================
%=============================================================================
%=============================================================================

%%\bibliographystyle{utphys_spires}
%% % \bibliographystyle{h-physrev3}
%%\bibliography{radcor}
\providecommand{\href}[2]{#2}
%=============================================================================
%=============================================================================
%=============================================================================
\end{document}